\documentclass[apjl, reprint, superscriptaddress, nofootinbib]{revtex4-1}

\usepackage{amsmath}
\usepackage{amssymb}
\usepackage{graphicx}
\usepackage{color}
\usepackage{subfig}
\usepackage{bm}
\usepackage{gensymb}
\usepackage{bigints}

\usepackage[usenames,dvipsnames,svgnames,table]{xcolor}
\usepackage{aas_macros}
\usepackage{physics}
\usepackage[sort&compress]{natbib}
\begin{document}

\preprint{APS/123-QED}

\title{A Topological Data Analysis of the CHIME/FRB Catalogues}

\author{Shruti Bhatporia}
 \email{shrutibhatporia@gmail.com}
\affiliation{High Energy Physics, Cosmology \& Astrophysics Theory (HEPCAT) Group, Department of Mathematics \& Applied Mathematics,
University of Cape Town, Cape Town 7700, South Africa}
 
\author{Anthony Walters}%
\email{waltersa@ukzn.ac.za}
\affiliation{Astrophysics Research Centre, University of KwaZulu-Natal, Westville Campus, Durban 4041, South Africa}
\affiliation{School of Mathematics, Statistics \& Computer Science, University of KwaZulu-Natal, Westville Campus, Durban 4041, South Africa}
\affiliation{High Energy Physics, Cosmology \& Astrophysics Theory (HEPCAT) Group, Department of Mathematics \& Applied Mathematics, University of Cape Town, Cape Town 7700, South Africa}

\author{Jeff Murugan}%
 \email{jeff.murugan@uct.ac.za}
\affiliation{The Laboratory for Quantum Gravity \& Strings, Department of Mathematics and Applied Mathematics, University of Cape Town}

\author{Amanda Weltman}%
 \email{amanda.weltman@uct.ac.za}
\affiliation{High Energy Physics, Cosmology \& Astrophysics Theory (HEPCAT) Group, Department of Mathematics \& Applied Mathematics, University of Cape Town, Cape Town 7700, South Africa}

\begin{abstract}
In this paper, we use Topological Data Analysis (TDA), a mathematical approach for studying data shape, to analyse Fast Radio Bursts (FRBs). Applying the Mapper algorithm, we visualise the topological structure of a large FRB sample. Our findings reveal three distinct FRB populations based on their inferred source properties, and show a robust structure indicating their morphology and energy. We also identify potential non-repeating FRBs that might become repeaters based on proximity in the Mapper graph. This work showcases TDA's promise in unraveling the origin and nature of FRBs.

\end{abstract}

%Use showkeys class option if keyword

                    %display desired
\maketitle

%\tableofcontents

\section{\label{sec:intro}Introduction}
FRBs are transient highly energetic pulses of radio waves that last only a few milliseconds. While one has been observed within our galaxy \cite{2022ApJ...926..121L}, the majority are extragalactic \cite{2007Sci...318..777L}, many now with identified host galaxies. The progenitor mechanism is still uncertain, though we expect them to be associated with extreme astrophysical phenomena such as neutron stars, black holes, and magnetars \cite{2019PhR...821....1P}. 
%FRBs are detected by radio telescopes, and arrays, that scan the sky for radio signals.  
A particularly perplexing property of FRBs is that a relatively small subset of observed bursts have been observed to repeat. To date, no periodic repeaters have been found and the timescales between repetitions are varied, thus it is difficult to prove that all FRBs are not repeaters.  This has led to speculation about two or more distinct populations of progenitors that give rise to these bursts \cite{2021ApJ...923....1P, 2022MNRAS.511.1961H, 2022JCAP...07..010G, 2022MNRAS.509.1227C, 2022MNRAS.509.1227C, 2022MNRAS.514.5987K}.

In this paper, we present a novel approach to analyse recent data from the Canadian Hydrogen Intensity Mapping Experiment (CHIME) collaboration on FRBs \cite{2018ApJ...863...48C}. CHIME is a radio telescope located in British Columbia, Canada, consisting of four cylindrical reflectors with 256 dual-polarization antennas each. Operating  in the frequency range of 400-800 MHz, it has a large field of view of about 200 square degrees.

High-dimensional data, as encountered, presents a formidable challenge in various scientific domains, including astrophysics. Its inherent complexity arises from the inability to visually grasp its entirety, especially when it is noisy and incomplete. As the number of parameters or dimensions in a dataset grows, the volume occupied by the data expands exponentially, and, in the real world, where observations are limited, high-dimensional datasets often become sparsely populated. This phenomenon, often referred to as the ``curse of dimensionality'', poses a significant obstacle to extracting meaningful insights and organising data efficiently, a challenge rarely encountered in lower-dimensional settings. In addressing this challenge, one critical strategy is dimensionality reduction (DR), which aims to uncover essential low-dimensional structures within high-dimensional datasets. These low-dimensional representations not only facilitate visualisation but also aid in the analysis and extraction of valuable information.

Effectively reducing dimensions can be transformative, shedding light on hidden patterns and relationships within the data. The most straightforward method of DR involves projection, which entails mapping the data onto lower-dimensional spaces. For instance, focusing on a subset of data features corresponds to a projection along specific axes. While this technique can reveal crucial insights, it may also obscure the true distances between neighbouring data points. Thus, the choice of projection is a pivotal step in the analysis. Among the most prevalent DR techniques is Principal Component Analysis (PCA). PCA identifies linear combinations of original features, known as principal components, that capture the maximum variance in the data. This method has been widely employed for dimensionality reduction and visualisation. It simplifies complex data while preserving much of the data's inherent structure. In addition to PCA, several sophisticated DR algorithms have emerged, addressing the limitations of linear projections. Nonlinear iterative approaches, such as t-Distributed Stochastic Neighbor Embedding (t-SNE), Uniform Manifold Approximation and Projection (UMAP), and the recent Pairwise Controlled Manifold Approximation Projection (PaCMAP) \cite{JMLR:v22:20-1061}, aim to uncover nonlinear relationships in the data. These methods excel at preserving the closeness of data points, yielding low-dimensional representations that are rich in meaning and facilitate advanced analyses.

Extending and updating the results first reported in \cite{2019arXiv190411044M}, we apply topological data analysis (TDA) to the recent CHIME/FRB dataset, with a focus on repeating FRBs. TDA is a branch of applied mathematics that uses techniques from topology - the study of the shape and connectivity of spaces - to infer relevant features within the data from complex and high-dimensional data sets. Specifically, it provides tools to capture qualitative properties of data that are invariant under deformations. As such, TDA can reveal hidden patterns, clusters, and structures in data that are robust to noise and insensitive to the choice of metric.

%We apply TDA to the CHIME/FRB dataset using the Mapper algorithm.
We use the first CHIME/FRB catalogue, which contains 535 FRBs detected by CHIME between 25 July 2018 and 1 July 2019, including 62 bursts from 18 repeating sources supplemented with a repeater catalogue containing 146 bursts of 25 repeaters detected during  from 2019 September 30
to 2021 May 1. We use various features of FRBs such as luminosity and energy of the burst based on fluence (the integrated flux over time), intrinsic width (the duration of the burst after correcting for dispersion and scattering), spectral bandwidth (the range of frequencies over which the burst is detected), and frequency slope (the change in frequency over time) to parameterise the data space. 

Our main goal is to use TDA to make predictions for where to find new repeaters among the FRBs detected by CHIME. We hypothesise that repeating FRBs have some distinctive topological features that can be captured by TDA. We aim to explore the similarities and differences between repeaters and non-repeaters in terms of their physical properties and their distribution in the sky. We hope that our analysis can shed some light on the nature and origin of FRBs and their possible applications to cosmology and fundamental physics.

%The rest of this article is organised as follows: In Section \ref{sec:tda} we introduce the basics of topological data analysis focusing on its use in clustering noisy data. Sections \ref{sec:frbs} and \ref{sec:data} describe the catalogued dataset that we use, motivates the need for intrinsic parameters and describes data-cuts and standardisation techniques. Finally, in Section \ref{sec:results} we discusses our results and analysis from the algorithm.

\section{\label{sec:tda}Topological Data Analysis}
Within the domain of TDA, Persistent Homology and the Mapper algorithm represent two major tools to explore the shape of a dataset, each with distinct foci and applications. A broad outline of how these two algorithms work is given below.

\subsection{\label{sec:ph}Persistent Homology}
Persistent homology operates on the fundamental concept that topological features, such as connected components (0-dimensional), loops (1-dimensional), or voids (2-dimensional), can persist across different levels of granularity. The algorithm focuses on capturing features that persist and evolve as the filtration parameter changes, making it particularly suitable for understanding the global topological characteristics of data. It proceeds as follows:

\begin{itemize}
    \item {\bf Filtration:} The process begins by constructing a simplicial complex from the data points. A simplicial complex is a mathematical representation of topological features. The filtration, a continuous parameter like distance or radius, gradually increases, considering data points in an ascending order of this parameter.
    \item {\bf Complex Evolution:} As the filtration parameter increases, new simplices (vertices, edges, triangles, etc.) are added to the complex. This process mimics the ``growing" of topological features.
    \item{\bf Tracking Birth and Death:} Persistent homology keeps track of when these topological features are ``born" and ``die" as the filtration parameter increases. A feature is born when it first appears in the complex, and it dies when it merges or disappears.
\end{itemize}

To represent this persistence of topological features, persistent homology uses visual tools like {\it persistence diagrams} or {\it barcodes}. A persistence diagram is a scatter plot with birth on the x-axis and death on the y-axis. Each point on the diagram corresponds to a topological feature (connected component, loop, void). The horizontal position of a point indicates when the feature is born, and the vertical position represents when it dies.

\subsection{\label{sec:mapper}The Mapper Algorithm}

The second major tool in the TDA framework is the Mapper algorithm that, like Persistent Homology, facilitates the exploration and representation of the shape and structure of complex data. Unlike Persistent Homology, which focuses on capturing topological features across different resolutions, Mapper offers a different approach that can be summarised as follows:
\begin{itemize}
    \item {\bf Dimensionality Reduction:} Mapper begins by mapping high-dimensional data to a lower-dimensional space using a filter function or lens. This filter function can be based on various techniques like projecting data onto principal components or estimating data density. The goal is to simplify the data while retaining its essential characteristics.
    \item {\bf Cover Construction:} In this step, Mapper constructs a cover of the lower-dimensional projected space. This cover is formed by dividing the space into overlapping intervals or hyper-balls of constant size. The choice of intervals or balls depends on the specific dataset and the problem being analysed.
    \item{\bf Clustering:} Within each interval or ball, Mapper applies a clustering algorithm of choice to group data points that are close to each other. Common clustering methods like k-means or DBSCAN can be used. This step aims to identify local patterns or clusters within the data.
    \item{\bf Simplicial Complex:} Finally, Mapper builds a simplicial complex, which is a network composed of nodes and edges. Each cluster from the previous step corresponds to a node in the complex, and edges are established between clusters that share some common data points. This connectivity reveals the relationships between clusters and captures the global and local structure of the data.
\end{itemize}

While each is useful in its own domain, Mapper's approach of overlapping bins or intervals ensures local connectivity in the representation of data. This allows for a detailed exploration of clusters and patterns at various scales, making it suitable for analysing complex datasets. Identifying separate disconnected clusters or regions in the mapper graph can reveal different populations or distinct structures within the dataset.

\section{\label{sec:frbs}Fast Radio Bursts and Their Intrinsic Properties}
Observable quantities associated with FRBs contain information about the nature of the source, effects associated with propagation to the observer, and the instrument used to detect them. Notable among FRB observables is the dispersion measure (DM), which we further use to calculate luminosity and energy of the burst. The DM of an FRB quantifies the cumulative effect of free electrons along the line of sight and can be expressed as $\mathrm{DM} \sim \int n_e\, dl$. This integral accounts for contributions from distinct regions containing free electrons. For an extragalactic FRB, the observed DM can be partitioned as follows:
\begin{align}
{\rm DM}_{\rm obs} = {\rm DM}_{\rm MW} + {\rm DM}_{\rm IGM} + \frac{{\rm DM}_{\rm host}}{(1+z)}\,.
\end{align}
Here, $\mathrm{DM}_{\rm MW}$ encompasses contributions from the Milky Way, $\mathrm{DM}_{\rm IGM}$ represents an influence of intervening gas in the intergalactic medium (IGM), $\mathrm{DM}_{\rm host}$ accounts for the FRB host galaxy and the immediate source environment, and $z$ is its redshift. Owing to the lumpy distribution of matter in the Universe, $\mathrm{DM}_{\rm IGM}$ depends upon both the distance to the FRB and its position on the sky. However, averaging $\mathrm{DM}_{\rm IGM}$ over all possible sight-lines yields an expression for the mean IGM contribution, in terms of the background cosmology. At redshifts below $z\sim1$ the DM-$z$ relation is well approximated by a linear function and can be expressed as \cite{2020Natur.581..391M}
\begin{equation}
z \sim \frac{\langle\mathrm{DM}_{\rm IGM}\rangle}{1200}~\mathrm {pc\,cm}^{-3}. \label{eq:redshift}
\end{equation}
This expression can provide a crude proxy for distance to the source. By neglecting the host galaxy contribution to the observed DM, and subtracting off the relatively well-modelled Milky Way contribution, one can use (\ref{eq:redshift}) to place an upper limit on the redshift of the FRB. This does not give accurate redshift measurements and overestimates the distances by not including host galaxy and foreground galaxy distributions, if any.

%Once the redshift of a source is known, observed frequencies (and bandwidths) can be converted to their rest-frame values using the standard expression,
%\begin{equation}
%     \nu_{\rm rest}=\nu_{\rm obs}(1+z),
%     \label{eq : nu_shifted}
%\end{equation}
%where $\nu_{\rm rest}$ is the rest-frame frequency, and $\nu_{\rm obs}$ is the observed frequency.

Another important observable associated with FRBs is their spectral flux density, $S_{\nu}$. This contains information on the energetics of the source, together with propagation and detection effects. By assuming isotropic emission, and the inverse square law, one can approximate the intrinsic spectral luminosity as \cite{zhang2018fast, 2022ApJ...926..206Z}
\begin{equation}
    L_p \sim 4\pi D_L^2(z) S_{\nu}\nu\,.
    \label{eq:luminosity}
\end{equation}
Similarly, the total radiative energy emitted by a source can be determined using \cite{zhang2018fast, 2022ApJ...926..206Z}
\begin{equation}
    E \sim \frac{4\pi D_L^2(z)}{(1+z)} F_{\nu} \nu\,.
    \label{eq:energy}
\end{equation}
In these equations, $D_L(z)$ represents the luminosity distance, while $S_{\nu}$ denotes the average flux values. Using equations (\ref{eq:luminosity}) - (\ref{eq:energy}) with the approximate redshifted given by Eqn. (\ref{eq:redshift}), produce crude upper limits on intrinsic source properties, since contributions from cosmic inhomogeneities and the FRB host galaxy are neglected. We use the central frequency of emission with reported peak frequency interchangeably for energy calculations. This assumption is valid as CHIME/FRBs have well-defined peak frequency for a Gaussian-like spectral shape. In addition, frequency cutoff described in Section \ref{sec:data} removes bursts that have central emission frequencies outside receiver band. These expressions will suffice for our purposes since TDA is known to perform well in the presence of noise, and we will be using an FRB catalogue which does not contain redshift information.

%These calculations enable us to establish upper limits on both energy and luminosities of the bursts. 

\section{\label{sec:data}Description of the Data}
%The initial release of the CHIME/FRB catalogue comprised 536 bursts (or sub-bursts in case of multiple burst reception), categorised into two groups: 474 bursts are considered apparent non-repeaters, while the remaining 62 bursts are associated with 18 repeating sources \cite{2021ApJS..257...59A}. 
For each CHIME/FRB catalogue burst, there are $\sim30$ observable parameters reported. Some of these, such as the DM, scattering time, spectral shape, bandwidth, flux, and pulse width, are closely related to astrophysical processes. In contrast, others, such as the Signal-to-Noise Ratio (SNR) and positions on the sky etc., are more biased to their detection at the instrument. Since we aim to construct a dataset which reflects the intrinsic properties of FRBs in their rest frame, uncontaminated by propagation and instrumental effects, we use only a subset of those provided in the catalogue and derive approximate rest-frame quantities for each burst. We use the reported extragalactic DM to approximate the redshift according to Equation (\ref{eq:redshift}), and with this, we compute the isotropic luminosity using Equation (\ref{eq:luminosity}) and isotropic energy using Equation (\ref{eq:energy}). In addition, we shift all measured frequencies (and bandwidths) to their rest-frame values. A summary of the parameters we include in our dataset is shown in Table \ref{tab:der_params}.
\begin{table*}
\begin{ruledtabular}
\begin{tabular}{ll}
\textbf{Intrinsic Burst Properties}&\textbf{Description}\\
\hline
\texttt{width\_fitb} & upper limit sub-pulse width by \texttt{fitburst}\\
\texttt{sp\_idx} & spectral index of the sub-burst\\
\texttt{sp\_run} & frequency dependency of spectral shape\\
\texttt{restframe\_peak\_freq} & peak frequency of the pulse in MHz at FWTM$^\ast$\\
\texttt{restframe\_bw} & bandwidth obtained from \texttt{high\_freq} a \texttt{low\_freq} at FWTM$^\ast$\\
\texttt{log\_E} & energy values of the burst\\
\texttt{log\_Lp} & luminosity of the the burst\\
\end{tabular}
\end{ruledtabular}
\caption{\label{tab:der_params} CHIME/FRB catalogue parameters included in this analysis, together with the derived ``rest frame'' parameters. ($^\ast$Full Width at a Tenth Maximum)} 
\end{table*}

To enhance data reliability and mitigate unwanted biases arising from instrumental effects, we apply further cuts to the catalogue. In particular, we exclude bursts with low SNR. We also exclude low-DM bursts from the sample, as these may be dominated by Milky Way subtraction and host galaxy uncertainties. And we exclude bursts with peak frequencies falling outside of the $400-800$ MHz band since we use these values are used to compute the rest-frame luminosity and energy. 
\begin{enumerate}
    \item $\mathrm{SNR} > 12$
    \item $\mathrm{DM} > 1.5 ~\mathrm{max}(DM_{\mathrm{NE2001}}, DM_{\mathrm{YMW16}})$
    \item $400.1 \leq \nu_{\rm peak} \leq 799.9$ MHz
\end{enumerate}
Once these cuts have been applied, we are left with 584 bursts in the sample, of which 185 are flagged as repeaters. To ensure accuracy, we utilise more precise \texttt{fitburst} values calculated offline for SNR, DM, and pulse width, as real-time measurements often come with larger error bars due to limited time resolutions within the receiver system.

Since the features in our dataset have vastly different ranges, we apply a prepossessing transformation to avoid the need for a non-Euclidean metric in the clustering algorithm employed by Mapper. Various transformations were explored. We found that most linear preprocessing methods yielded comparable clustering outcomes for repeaters and non-repeaters, and opted for \texttt{StandardScalar}, which scales all data features such that the have zero mean and unit variance. For further detail on the choice of prepossessing transformation, see Appendix ...

\section{\label{sec:results}Results}
We apply the TDA methods described in \S\ref{sec:tda} to the FRB dataset described in \S\ref{sec:data}, using the Python packages \texttt{Ripser} \cite{ctralie2018ripser}, and \texttt{kepler-mapper} \cite{KeplerMapper_JOSS}. 

\begin{figure*}[htp!]
    \centering
    \includegraphics[scale=1.08]{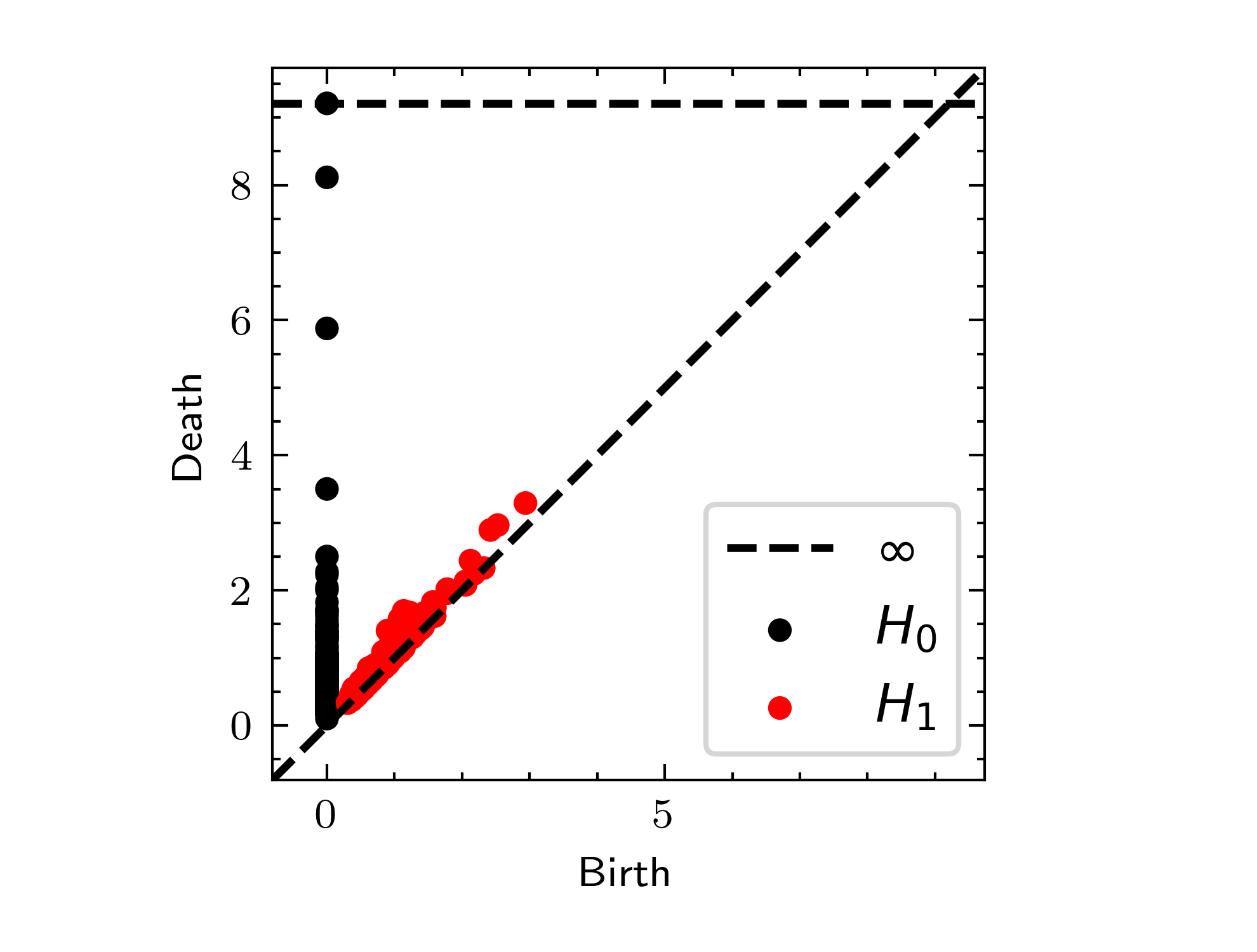}
    \includegraphics[scale=0.4]{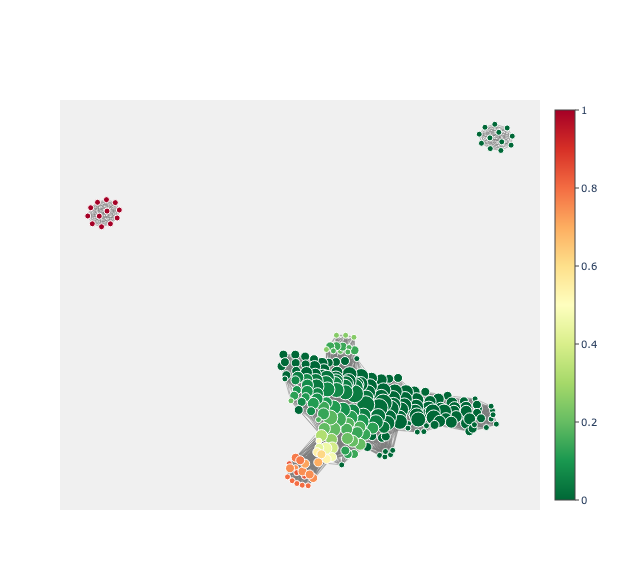}
    \caption{The persistence diagram (left) and Mapper output (right) for the rest-frame FRB datasets described in \S\ref{sec:data}. The persistence diagram shows the birth and death of the first two homology groups, $H_0$ (black) and $H_1$ (red). The mapper output depicts clusters of repeaters and apparent non-repeaters. The size of each node is proportional to the number of FRBs it contains, coloured-coded according to the faction of repeaters. }
    \label{fig:tda}
\end{figure*}

The persistence diagram shown in left panel of Figure \ref{fig:tda} indicates the birth and death of the first two homology groups, $H_0$ (black) and $H_1$ (red). Note in particular the two $H_0$ points which persist over a large range of scales (birth 0, death $>$8) and are associated with the three groupings in Figure 2. This lends support to our observation about these distinct groups, since the mapper result is quite sensitive to choice of hyperparameters (scale), whereas persistent homology tells us about structures across a range of scales. Note also the absence of any persistent $H_1$ groups in our data which indicates an absence of any loops in the data.

As demonstrated in Figure \ref{fig:mapper_obs_param_pca} in Appendix \ref{sec:appA}, clustered groups exhibit varying characteristics based on different parameters. Specifically, separate clusters are observed to have low spectral running and low bandwidth. Conversely, repeaters are discerned by higher spectral indices, narrower bandwidth, and lower SNR. We find that there exists a subgroup among repeaters characterised by high SNR but relatively lower luminosity and spectral running.

Our analysis identifies apparent non-repeaters that potentially share characteristics with repeaters. These candidates are situated in nodes where more than half of the elements are repeaters. We designate such a node as a ``repeater node''. For instance, we find that single-burst candidates grouped with higher SNR values tend to exhibit higher energy levels and are more amenable to follow-up observations, thus falling into repeater nodes. The repeater nodes are marked with warm red-toned colours in the right panel of Figure \ref{fig:tda}.

To gain further insights, we construct a persistence homology diagram described in Section \ref{sec:ph} to detect any topological features in the data. For the CHIME/FRB parameters considered in this study, the diagram in left panel of Figure \ref{fig:tda}. reveals three persistent features that confirm three clusters in the mapper analysis. 
Further analysis uncovers that among the repeater candidates, FRB20190430C was initially identified as single-burst FRBs but were later found to exhibit repeating behaviour in the CHIME/FRB 2023 catalogue \cite{2021ApJS..257...59C}. Based on our simulations, we identify FRB20181221A, along with the candidates listed in Table \ref{tab:candidates}, as intriguing candidates warranting further investigation for repeater characteristics. To validate our findings, we confirm the clustering patterns of repeaters and non-repeaters within the mapper by projecting known repeater candidates back into two-dimensional PCA projections. This analysis effectively corroborates that all repeater candidates exhibit similar properties to those of known repeaters, as evidenced in Figure \ref{fig:pca_proj_cand}. Proposed repeater candidates in the figure are marked in blue, with red denoting known repeaters and black are apparent non-repeaters.

\begin{table}[htp!]
\begin{ruledtabular}
\begin{tabular}{ll}

    \textbf{Lens/Filter} & \textbf{Repeater Candidates}\\
    \hline
    PCA & FRB20181129B, FRB20181017B,\\ & FRB20181213B, FRB20190112A, \\& FRB20190218B, FRB20190228A, \\
    & FRB20190228A, FRB20190403G, \\ & FRB20190422A, FRB20181128C, \\ & FRB20181203B, FRB20190129A, \\
    & FRB20190621C
\end{tabular}
\end{ruledtabular}
\caption{\label{tab:candidates}Repeaters candidates from different algorithms }
\end{table}

\section{\label{sec:discussion}Summary}
In this paper, we have applied topological data analysis (TDA) to study the properties of fast radio bursts (FRBs) detected by the CHIME/FRB experiment. We have used the mapper algorithm to construct networks that capture global shape of parameter space. We have focused on the repeater and non-repeater populations. Our main findings are:

\begin{itemize}
    \item The sources of repeaters and non-repeaters are uniformly distributed over the sky, indicating their extragalactic origin and even distribution for low redshifts.
    \item The bursts from three distinct clusters in the mapper networks are based on their inferred source properties. One of these clusters contains the majority of, and is exclusively associated with, repeating bursts.
    \item The repeater population can be further divided into three sub-groups based on clustering shown in Figure \ref{fig:tda} and Figure \ref{fig:mapper_obs_param_pca}. One subgroup is Based on their spectral properties ie spectral index and spectral running and bandwidth, luminosity and energy these sub-groups may hint different types of mechanisms that produce repeating FRBs.
    \item The PCA projections and mappers do not reveal any clear trend or dichotomy among the repeaters themselves, therefore we conclude the repeater properties do not seem to depend on the given intrinsic parameter space in the 2023 catalogue.
\end{itemize}

Most repeaters in our dataset exhibit larger \texttt{bc\_width} and narrower \texttt{bw}. If these repeaters share similarities with FRB121102 and are situated within a dense circumgalactic environment, they should have a large \texttt{scat\_time}, or a significant portion of expected repeaters should display substantial scattering times.\\

We conclude that TDA is a powerful tool to explore the diversity and complexity of the FRB phenomenon. It can help us to identify and classify different populations and sub-populations of FRBs, and to reveal their intrinsic and extrinsic features. We suggest that future studies should apply TDA to larger and more complete FRB samples, and incorporate other parameters such as polarisation, dispersion measure and rotation measure. We recommend that follow-up observations of the repeater candidates should be conducted to verify their nature and origin.

\acknowledgments

We thank Mugundhan Vijayraghavan and Surajit Kalita for their inputs on parameter selections and datacuts. S.B. thanks NRF post-graduate grant number MND200622534919 for funding the research. We gratefully acknowledge support from the University of Cape Town Vice Chancellor’s Future Leaders 2030 Awards programme which has generously funded this research and support from the South African Research Chairs Initiative of the Department of Science and Technology and the National Research Foundation. J.M and A.W. would like to acknowledge support from the ICTP through the Associates Programme and from the Simons Foundation through grant number 284558FY19.\\

%\newpage
\bibliographystyle{unsrt}
\bibliography{frbtda_bib}% Produces the bibliography via BibTeX.

\begin{thebibliography}{10}

\bibitem{2022ApJ...926..121L}
J.~D. {Lyman}, A.~J. {Levan}, K.~{Wiersema}, C.~{Kouveliotou}, A.~A. {Chrimes},
  and A.~S. {Fruchter}.
\newblock {The Fast Radio Burst-emitting Magnetar SGR 1935+2154-Proper Motion
  and Variability from Long-term Hubble Space Telescope Monitoring}.
\newblock {\em \apj}, 926(2):121, February 2022.

\bibitem{2007Sci...318..777L}
D.~R. {Lorimer}, M.~{Bailes}, M.~A. {McLaughlin}, D.~J. {Narkevic}, and
  F.~{Crawford}.
\newblock {A Bright Millisecond Radio Burst of Extragalactic Origin}.
\newblock {\em Science}, 318(5851):777, November 2007.

\bibitem{2019PhR...821....1P}
E.~{Platts}, A.~{Weltman}, A.~{Walters}, S.~P. {Tendulkar}, J.~E.~B. {Gordin},
  and S.~{Kandhai}.
\newblock {A living theory catalogue for fast radio bursts}.
\newblock {\em \physrep}, 821:1--27, August 2019.

\bibitem{2021ApJ...923....1P}
Ziggy {Pleunis}, Deborah~C. {Good}, Victoria~M. {Kaspi}, Ryan {Mckinven},
  Scott~M. {Ransom}, Paul {Scholz}, Kevin {Bandura}, Mohit {Bhardwaj}, P.~J.
  {Boyle}, Charanjot {Brar}, Tomas {Cassanelli}, Pragya {Chawla}, Fengqiu
  {(Adam) Dong}, Emmanuel {Fonseca}, B.~M. {Gaensler}, Alexander {Josephy},
  Jane~F. {Kaczmarek}, Calvin {Leung}, Hsiu-Hsien {Lin}, Kiyoshi~W. {Masui},
  Juan {Mena-Parra}, Daniele {Michilli}, Cherry {Ng}, Chitrang {Patel}, Masoud
  {Rafiei-Ravandi}, Mubdi {Rahman}, Pranav {Sanghavi}, Kaitlyn {Shin},
  Kendrick~M. {Smith}, Ingrid~H. {Stairs}, and Shriharsh~P. {Tendulkar}.
\newblock {Fast Radio Burst Morphology in the First CHIME/FRB Catalog}.
\newblock {\em \apj}, 923(1):1, December 2021.

\bibitem{2022MNRAS.511.1961H}
Tetsuya {Hashimoto}, Tomotsugu {Goto}, Bo~Han {Chen}, Simon C.~C. {Ho}, Tiger
  Y.~Y. {Hsiao}, Yi~Hang~Valerie {Wong}, Alvina Y.~L. {On}, Seong~Jin {Kim},
  Ece {Kilerci-Eser}, Kai-Chun {Huang}, Daryl Joe~D. {Santos}, and Shotaro
  {Yamasaki}.
\newblock {Energy functions of fast radio bursts derived from the first
  CHIME/FRB catalogue}.
\newblock {\em \mnras}, 511(2):1961--1976, April 2022.

\bibitem{2022JCAP...07..010G}
Han-Yue {Guo} and Hao {Wei}.
\newblock {A possible subclassification of fast radio bursts}.
\newblock {\em \jcap}, 2022(7):010, July 2022.

\bibitem{2022MNRAS.509.1227C}
Bo~Han {Chen}, Tetsuya {Hashimoto}, Tomotsugu {Goto}, Seong~Jin {Kim}, Daryl
  Joe~D. {Santos}, Alvina Y.~L. {On}, Ting-Yi {Lu}, and Tiger Y.~Y. {Hsiao}.
\newblock {Uncloaking hidden repeating fast radio bursts with unsupervised
  machine learning}.
\newblock {\em \mnras}, 509(1):1227--1236, January 2022.

\bibitem{2022MNRAS.514.5987K}
Seong~Jin {Kim}, Tetsuya {Hashimoto}, Bo~Han {Chen}, Tomotsugu {Goto}, Simon
  C.~C. {Ho}, Tiger Yu-Yang {Hsiao}, Yi~Hang~Valerie {Wong}, and Shotaro
  {Yamasaki}.
\newblock {On the relationship between the duration and energy of non-repeating
  fast radio bursts: census with the CHIME data}.
\newblock {\em \mnras}, 514(4):5987--5995, August 2022.

\bibitem{2018ApJ...863...48C}
{CHIME/FRB Collaboration}, M.~{Amiri}, K.~{Bandura}, P.~{Berger},
  M.~{Bhardwaj}, M.~M. {Boyce}, P.~J. {Boyle}, C.~{Brar}, M.~{Burhanpurkar},
  P.~{Chawla}, J.~{Chowdhury}, J.~F. {Cliche}, M.~D. {Cranmer}, D.~{Cubranic},
  M.~{Deng}, N.~{Denman}, M.~{Dobbs}, M.~{Fandino}, E.~{Fonseca}, B.~M.
  {Gaensler}, U.~{Giri}, A.~J. {Gilbert}, D.~C. {Good}, S.~{Guliani},
  M.~{Halpern}, G.~{Hinshaw}, C.~{H{\"o}fer}, A.~{Josephy}, V.~M. {Kaspi},
  T.~L. {Landecker}, D.~{Lang}, H.~{Liao}, K.~W. {Masui}, J.~{Mena-Parra},
  A.~{Naidu}, L.~B. {Newburgh}, C.~{Ng}, C.~{Patel}, U.~L. {Pen},
  T.~{Pinsonneault-Marotte}, Z.~{Pleunis}, M.~{Rafiei Ravandi}, S.~M. {Ransom},
  A.~{Renard}, P.~{Scholz}, K.~{Sigurdson}, S.~R. {Siegel}, K.~M. {Smith},
  I.~H. {Stairs}, S.~P. {Tendulkar}, K.~{Vanderlinde}, and D.~V. {Wiebe}.
\newblock {The CHIME Fast Radio Burst Project: System Overview}.
\newblock {\em \apj}, 863(1):48, August 2018.

\bibitem{JMLR:v22:20-1061}
Yingfan Wang, Haiyang Huang, Cynthia Rudin, and Yaron Shaposhnik.
\newblock Understanding how dimension reduction tools work: An empirical
  approach to deciphering t-sne, umap, trimap, and pacmap for data
  visualization.
\newblock {\em Journal of Machine Learning Research}, 22(201):1--73, 2021.

\bibitem{2019arXiv190411044M}
Jeff {Murugan} and Duncan {Robertson}.
\newblock {An Introduction to Topological Data Analysis for Physicists: From
  LGM to FRBs}.
\newblock {\em arXiv e-prints}, page arXiv:1904.11044, April 2019.

\bibitem{2020Natur.581..391M}
J.~P. {Macquart}, J.~X. {Prochaska}, M.~{McQuinn}, K.~W. {Bannister},
  S.~{Bhandari}, C.~K. {Day}, A.~T. {Deller}, R.~D. {Ekers}, C.~W. {James},
  L.~{Marnoch}, S.~{Os{\l}owski}, C.~{Phillips}, S.~D. {Ryder}, D.~R. {Scott},
  R.~M. {Shannon}, and N.~{Tejos}.
\newblock {A census of baryons in the Universe from localized fast radio
  bursts}.
\newblock {\em \nat}, 581(7809):391--395, May 2020.

\bibitem{zhang2018fast}
Bing Zhang.
\newblock Fast radio burst energetics and detectability from high redshifts.
\newblock {\em The Astrophysical Journal Letters}, 867(2):L21, 2018.

\bibitem{2022ApJ...926..206Z}
Shu-Qing {Zhong}, Wen-Jin {Xie}, Can-Min {Deng}, Long {Li}, Zi-Gao {Dai}, and
  Hai-Ming {Zhang}.
\newblock {Can a Single Population Account for the Discriminant Properties in
  Fast Radio Bursts?}
\newblock {\em \apj}, 926(2):206, February 2022.

\bibitem{ctralie2018ripser}
Christopher Tralie, Nathaniel Saul, and Rann Bar-On.
\newblock {Ripser.py}: A lean persistent homology library for python.
\newblock {\em The Journal of Open Source Software}, 3(29):925, Sep 2018.

\bibitem{KeplerMapper_JOSS}
Hendrik~Jacob van Veen, Nathaniel Saul, David Eargle, and Sam~W. Mangham.
\newblock Kepler mapper: A flexible python implementation of the mapper
  algorithm.
\newblock {\em Journal of Open Source Software}, 4(42):1315, 2019.

\bibitem{2021ApJS..257...59C}
{CHIME/FRB Collaboration}, Mandana {Amiri}, Bridget~C. {Andersen}, Kevin
  {Bandura}, Sabrina {Berger}, Mohit {Bhardwaj}, Michelle~M. {Boyce}, P.~J.
  {Boyle}, Charanjot {Brar}, Daniela {Breitman}, Tomas {Cassanelli}, Pragya
  {Chawla}, Tianyue {Chen}, J.~F. {Cliche}, Amanda {Cook}, Davor {Cubranic},
  Alice~P. {Curtin}, Meiling {Deng}, Matt {Dobbs}, Fengqiu~Adam {Dong},
  Gwendolyn {Eadie}, Mateus {Fandino}, Emmanuel {Fonseca}, B.~M. {Gaensler},
  Utkarsh {Giri}, Deborah~C. {Good}, Mark {Halpern}, Alex~S. {Hill}, Gary
  {Hinshaw}, Alexander {Josephy}, Jane~F. {Kaczmarek}, Zarif {Kader}, Joseph~W.
  {Kania}, Victoria~M. {Kaspi}, T.~L. {Landecker}, Dustin {Lang}, Calvin
  {Leung}, Dongzi {Li}, Hsiu-Hsien {Lin}, Kiyoshi~W. {Masui}, Ryan {McKinven},
  Juan {Mena-Parra}, Marcus {Merryfield}, Bradley~W. {Meyers}, Daniele
  {Michilli}, Nikola {Milutinovic}, Arash {Mirhosseini}, Moritz
  {M{\"u}nchmeyer}, Arun {Naidu}, Laura {Newburgh}, Cherry {Ng}, Chitrang
  {Patel}, Ue-Li {Pen}, Emily {Petroff}, Tristan {Pinsonneault-Marotte}, Ziggy
  {Pleunis}, Masoud {Rafiei-Ravandi}, Mubdi {Rahman}, Scott~M. {Ransom}, Andre
  {Renard}, Pranav {Sanghavi}, Paul {Scholz}, J.~Richard {Shaw}, Kaitlyn
  {Shin}, Seth~R. {Siegel}, Andrew~E. {Sikora}, Saurabh {Singh}, Kendrick~M.
  {Smith}, Ingrid {Stairs}, Chia~Min {Tan}, S.~P. {Tendulkar}, Keith
  {Vanderlinde}, Haochen {Wang}, Dallas {Wulf}, and A.~V. {Zwaniga}.
\newblock {The First CHIME/FRB Fast Radio Burst Catalog}.
\newblock {\em \apjs}, 257(2):59, December 2021.

\end{thebibliography}

\appendix

\section{\label{sec:appA}Mapper plots with observed parameters}

\begin{figure*}[htp!]
    \centering
    \includegraphics[scale=0.5]{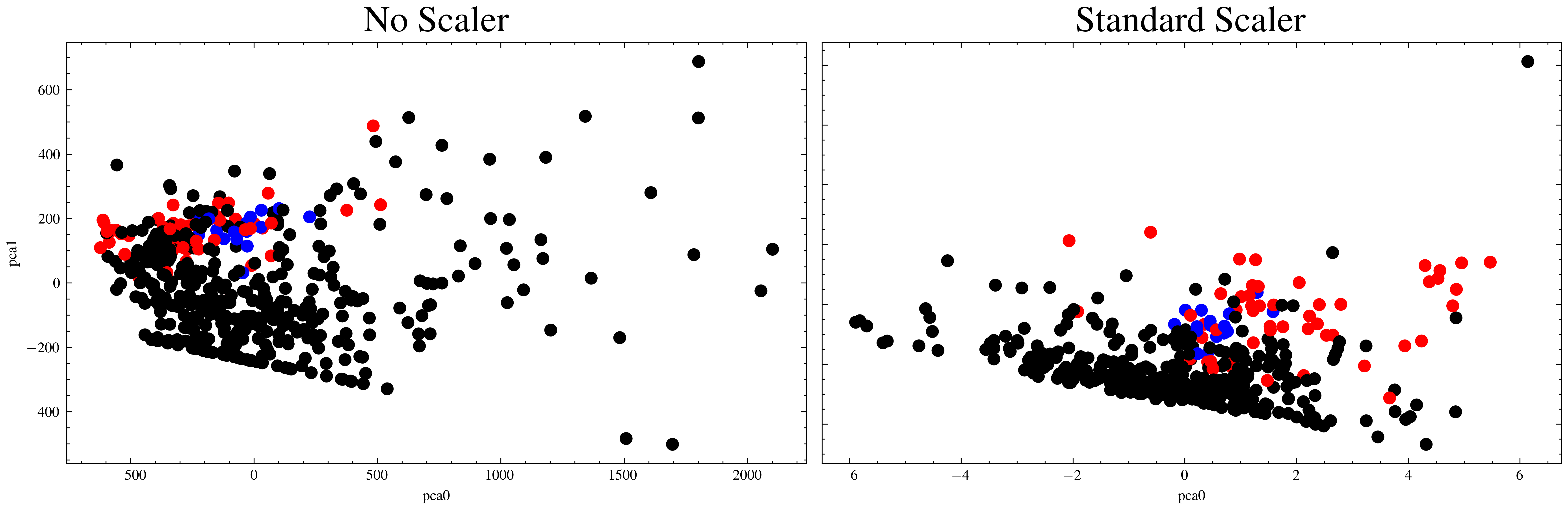}
    \caption{PCA projection of the data described in \S\ref{sec:data}, for two preprocessing transformations. Once-off bursts are shown in black, repeaters in red, and repeater candidates in blue.}
    \label{fig:pca_proj_cand}
\end{figure*}

To ensure the robustness of our Mapper generation process and prevent numerical extremes, we employ the standardscaler transformation shown in Figure \ref{fig:pca_proj_cand}. This transformation effectively centres the data, giving it a mean of zero and a standard deviation of one. In our evaluation, we have compared standardscaler with various other preprocessing techniques, aiming to assess their impact on clustering results. Notably, we observed that most linear preprocessing methods yielded comparable clustering outcomes for repeaters and non-repeaters.\\
As illustrated in Figure \ref{fig:pca_proj_cand}, this assessment of preprocessing techniques was conducted on a seven-dimensional dataset that was subsequently projected onto a two-dimensional space. It is worth highlighting that linear scaling algorithms such as standardscaler introduce a discernible separation between a repeater and apparent non-repeater clusters.\\

As in our detailed analysis in Section \ref{sec:results}, in this appendix, we unveil the intricate behaviours of crucial observational parameters within the identified clusters. These parameters are dispersion measure, spectral shape metrics including spectral index and spectral running, bandwidth, pulse width, and scattering time within the context of clustered phenomena. Notably, our examination uncovers discernible variations in luminosity and energy magnitudes within clusters of repeaters, shedding valuable light on the complex dynamics and characteristics of these intriguing astrophysical entities. \\

\begin{figure*}
\centering
\subfloat{\label{fig:pca_mapper_si}\includegraphics[width=0.26\textwidth]{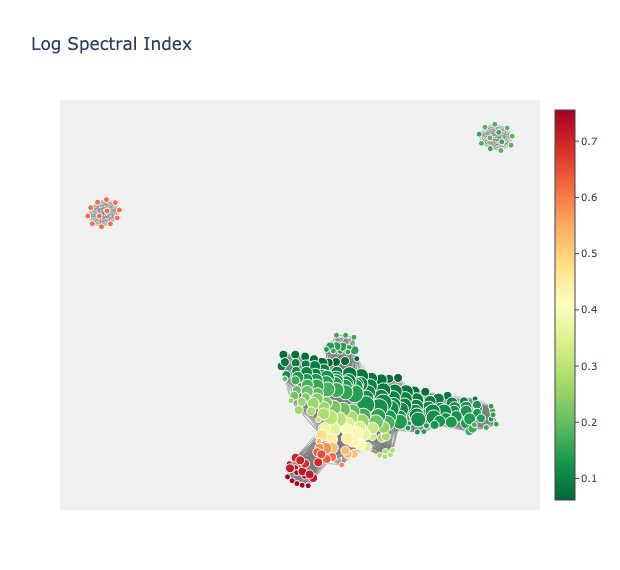}}\qquad
\subfloat{\label{fig:pca_mapper_bw}\includegraphics[width=0.26\textwidth]{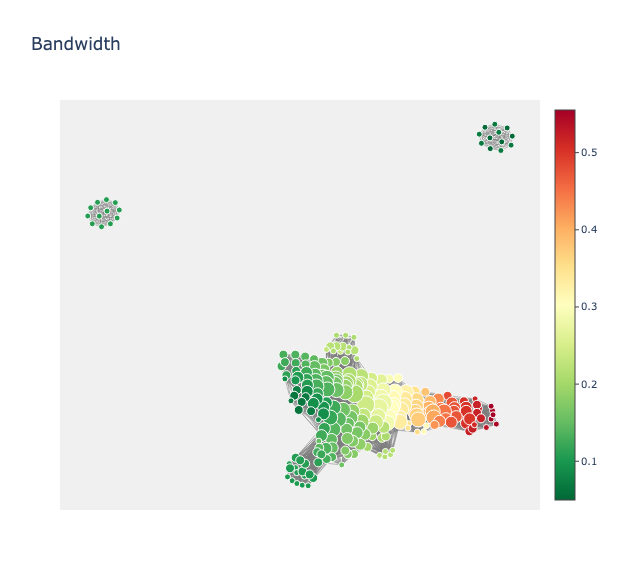}}\qquad
\subfloat{\label{fig:pca_mapper_sr}\includegraphics[width=0.26\textwidth]{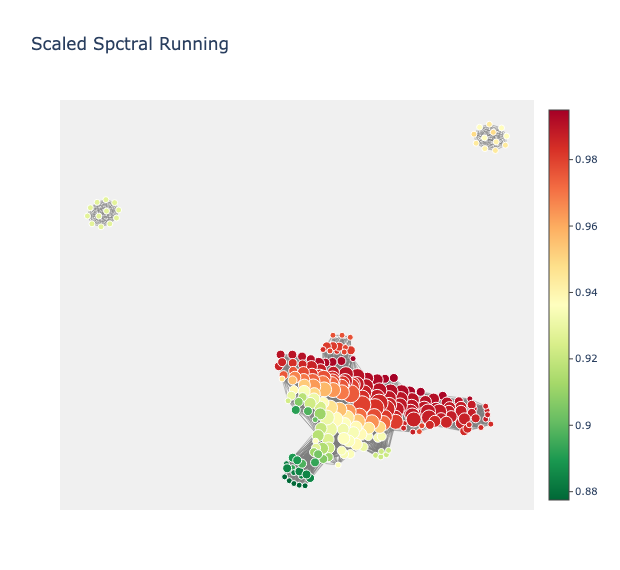}}\\
\subfloat{\label{fig:pca_mapper_snr}\includegraphics[width=0.26\textwidth]{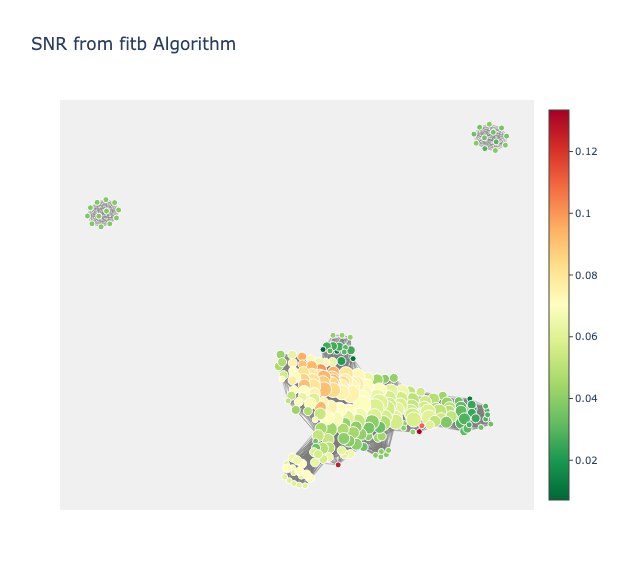}}\qquad
\subfloat{\label{fig:pca_mapper_bxw}\includegraphics[width=0.26\textwidth]{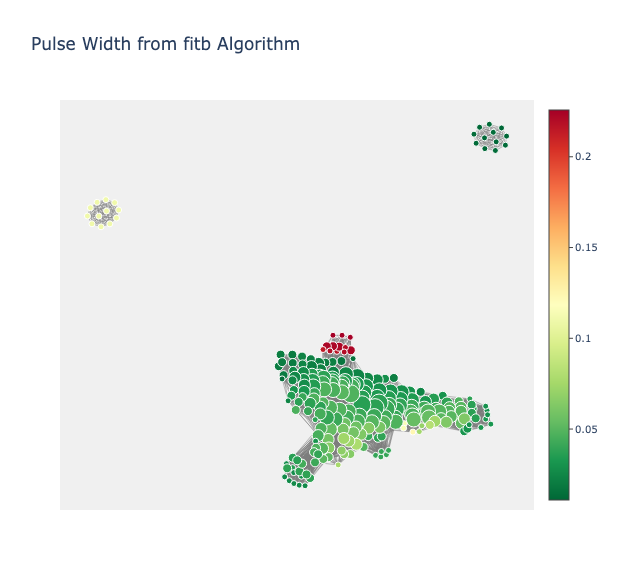}}\qquad
\subfloat{\label{fig:pca_mapper_lnlp}\includegraphics[width=0.26\textwidth]{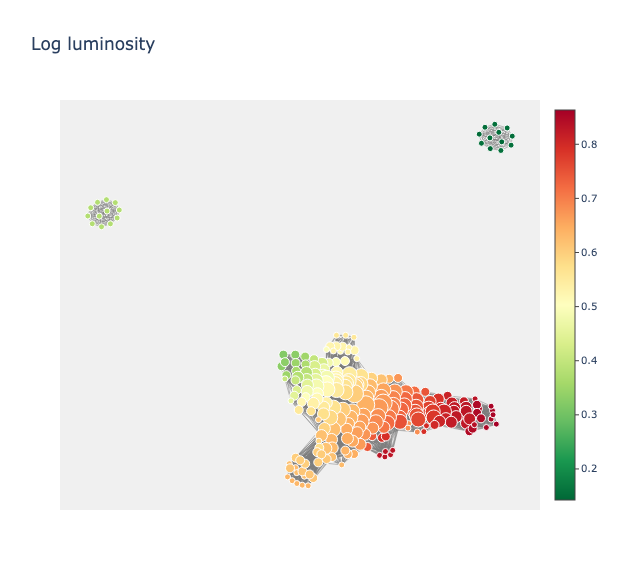}}\\
\subfloat{\label{fig:pca_mapper_lne}\includegraphics[width=0.26\textwidth]{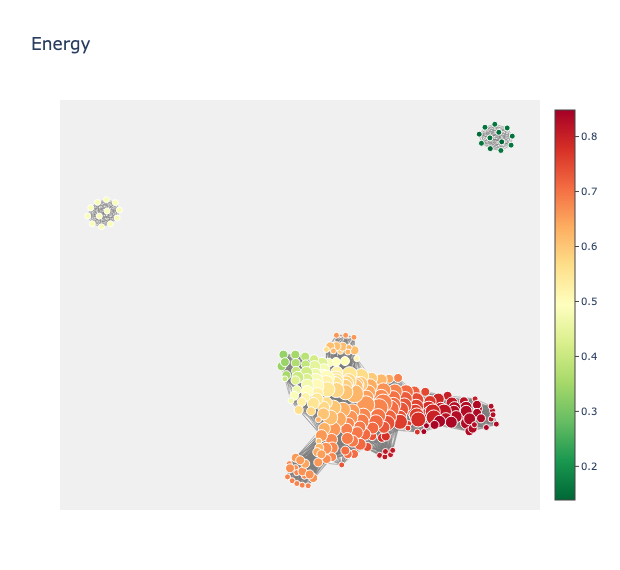}}\qquad
\subfloat{\label{fig:pca_mapper_dm}\includegraphics[width=0.26\textwidth]{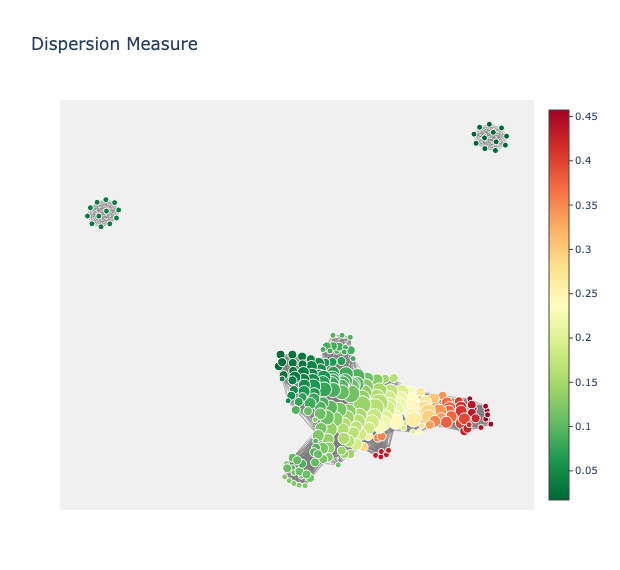}}\qquad
\subfloat{\label{fig:pca_mapper_sct}\includegraphics[width=0.26\textwidth]{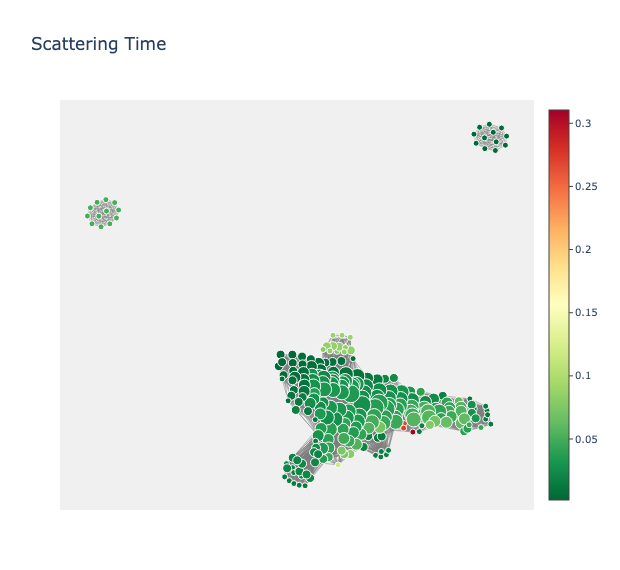}}\\
\caption{Mapper output coloured by catalogue parameter values}
\label{fig:mapper_obs_param_pca}
\end{figure*}

\end{document}